\documentclass[prd,nofootinbib,superscriptaddress,showpacs,twocolumn,reprint]{revtex4}
\usepackage[a4paper, total={6.5in, 8.5in}]{geometry}
\usepackage{graphicx}
\usepackage{bm}
\usepackage[usenames,dvipsnames]{color}
\usepackage{amsmath,amssymb}
\usepackage{slashed}
\usepackage{float}
\usepackage{epsfig}
\usepackage{comment}
\usepackage{amsfonts}
\usepackage{amssymb}
\usepackage{graphicx}
\usepackage{amsmath}
\usepackage{latexsym, amssymb} 
\usepackage[colorlinks=true,linkcolor=blue,citecolor=blue]{hyperref}
\usepackage{IEEEtrantools}
\usepackage{subfigure}
\begin{document}
\title{Gravitation Wave signal from Asteroid mass Primordial Black Hole Dark Matter} 
\author{Diptimoy Ghosh} 
\email{diptimoy.ghosh@iiserpune.ac.in}
\affiliation{Indian Institute of Science Education and Research, Pune 411008, India}
\author{Arvind Kumar Mishra} 
\email{arvind.mishra@acads.iiserpune.ac.in}
\affiliation{Indian Institute of Science Education and Research, Pune 411008, India}
\date{\today}

\begin{abstract}
Primordial Black Holes (PBHs) in the mass range $\sim 10^{17}- 10^{22}$g  are currently unconstrained, and can constitute the full Dark Matter (DM)
density of the universe. Motivated by this, in the current work, we aim to relate the existence of PBHs in the said mass range to the production
of observable Gravitational Waves (GWs) in the upcoming GW detectors. 
We follow a relatively model-independent approach assuming that the PBHs took birth in a radiation dominated era from enhanced
primordial curvature perturbation at small scales produced by inflation. We show that the constraints from CMB and BAO data allow
for the possibility of PBHs being the whole of DM density of the universe. Finally, we derive the GW spectrum induced by the enhanced
curvature perturbations  and show that they are detectable in the future GW detectors like eLISA, BBO and DECIGO.
\end{abstract}

\maketitle
\newpage

\section{Introduction}
\label{Sec1}
Primordial Black Holes (PBHs) are black holes which might have formed during the very early stages of the Universe \cite{zeldo:1966,Hawking:1971ei,Carr:1974nx,Carr:1975qj}. 
There are various proposed mechanisms for the formation of these PBHs, and their masses can span a very large range of values, see
\cite{Carr:2020xqk, Green:2020jor, Carr:2021bzv,Bird:2022wvk} for recent reviews .
Depending on the mass, PBHs can lead to different astrophysical and cosmological signatures which can be used
to discover/constrain their existence. The PBHs have not been detected yet, but if they exist, they may contribute to
Dark Matter (DM). For example, PBHs of mass $M_{\mathrm{PBH}}\leq 5\times 10^{14}$g are expected to have evaporated through Hawking radiation by now, and they cannot contribute to the Dark Matter density of the universe. Consequently, they do not produce astrophysical signatures, but may have effects on  Big Bang Nucleosynthesis (BBN) and/or
Cosmic Microwave Background (CMB) (see, for example, \cite{1977SvAL....3...76V,1977SvAL....3..110Z,Kohri:1999ex,Carr:2009jm,Acharya:2020jbv,Chluba:2020oip}).
Heavier PBHs in the mass range $5\times 10^{14}$g $\leq M_{\mathrm{PBH}}\leq 10^{17}$g are evaporating at present and can be constrained from non-observation of Hawking emission products \cite{Boudaud:2018hqb,Laha:2019ssq,DeRocco:2019fjq,Belotsky:2014twa,Chan:2020zry,Laha:2020ivk}. PBHs of mass more than $10^{22}$g are constrained by gravitational lensing and other considerations, see for example, \cite{Carr:2020gox, Carr:2020xqk, Green:2020jor, Carr:2021bzv} and the references therein. 
However, PBHs in the intermediate mass range $10^{17}- 10^{22}$g (often refereed to as asteroid-mass PBHs) are very poorly constrained
(for various efforts, see \cite{Capela:2012jz,Capela:2013yf,Pani:2014rca,Dasgupta:2019cae,Mittal:2021egv,Ghosh:2021gfa,Katz:2018zrn,Montero-Camacho:2019jte} ). 
Therefore, in this mass range, PBHs can constitute the whole of DM, see Fig. \cite{Carr:2020xqk}.
\begin{figure}[]
	\centering
\includegraphics[height=2.in,width=3.in]{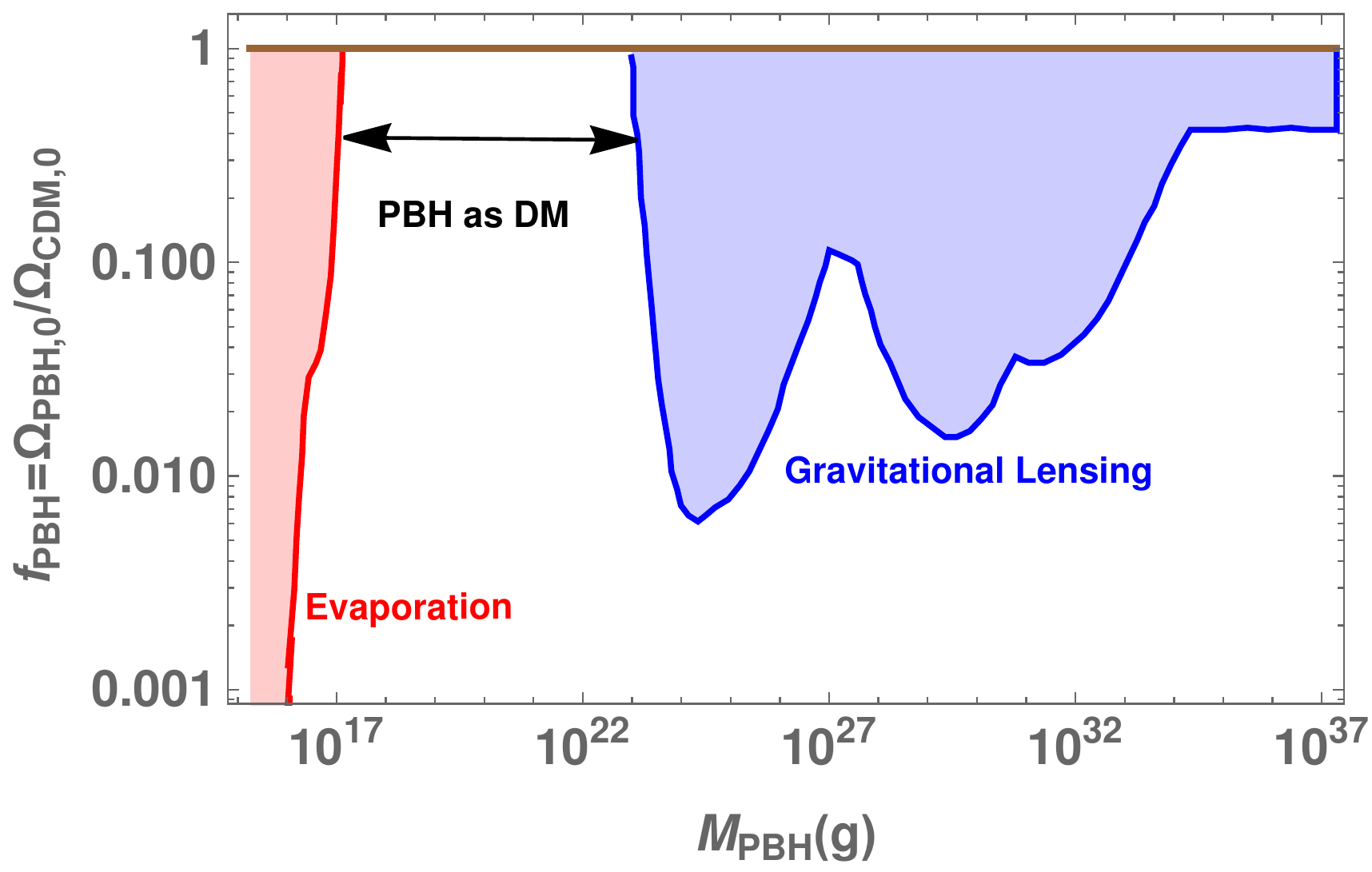}
	\label{fig:PBHfractmot}	
\caption{The observational constraints on the PBH DM fraction, $f_{\mathrm{PBH}}$, for a monochromatic mass function for the PBH
as a function of the PBH mass.
The constraints are taken from \cite{Carr:2020xqk}. }
	\label{fig:PBHDM}	
 \end{figure}

PBHs can form due to enhanced primordial curvature perturbation at small scales produced by inflation in the very early Universe \cite{zeldo:1966,Hawking:1971ei, Carr:1974nx,Hawking:1982ga,Hogan:1984zb, Hawking:1987bn}.
Since the scalar perturbations can give rise to also tensor modes at second order in perturbation theory,
PBH production in this case inevitably leads to the Scalar Induced Gravitational Wave (SIGW) generation \cite{Mollerach:2003nq,Ananda:2006af,Baumann:2007zm,Saito:2008jc,2010PhRvD..81b3517B,Alabidi:2012ex,Inomata:2018cht,Orlofsky:2016vbd,Ahmed:2021ucx,Kohri:2018awv,Fu:2019vqc,Gao:2021vxb,Yi:2020cut,Lin:2021vwc,Bhaumik:2020dor,Ragavendra:2020sop,Lin:2020goi}. We estimate the SIGW spectrum corresponding to PBH DM mass in the range $\sim 10^{17}- 10^{22}$g, and discuss the GW detection possibility
using the upcoming GW detectors like eLISA, BBO and DECIGO. Note that, in the literature, PBH formation in the mass range of our interest has been studied in the context of specific inflationary models, for example, see Refs. \cite{Braglia:2020eai,Spanos:2021hpk, Ahmed:2021ucx,Correa:2022ngq}. 
In our work, we do not consider any specific inflationary model, but  assume enhanced curvature power spectrum at small scales, see sec~\ref{Sec2}
for more details.

Our paper is organised as follows: In Section \ref{Sec1}, we briefly review the PBH formation, and associated secondary gravitational waves from enhanced scalar curvature perturbation. Section \ref{Sec2} describes the specific parametrisation of power spectrum adopted for our analysis. Our main results are
presented in Section \ref{Sec3}. Finally, we summarise our findings in Section \ref{Sec5}.

\section{PBH formation and SIGW generation}
\label{Sec1}
In this Section, we briefly review the primordial black hole formation and scalar induced gravitational wave generation. For a more detailed exposure
see, Refs. \cite{Carr:2020xqk, Green:2020jor, Carr:2021bzv,Bird:2022wvk,Ananda:2006af,Baumann:2007zm,Kohri:2018awv,Arya:2019wck,Arya:2022xzc}.
\subsection{Primordial Black Hole Formation}
\label{Subsec1}
As mentioned above, PBH formation can happen through various mechanisms, but here we focus on PBH generation via collapse
of primordial density perturbation (produced during inflation) in a radiation-dominated era.  In this mechanism, the PBHs form
when a overdensity, $\delta$, generated during inflation becomes larger than a critical density $\delta_{c}$, re-enters
into the horizon and collapses through gravitational instability \cite{Sasaki:2018dmp}. The mass of the PBHs produced during the time
when modes re-enter into the horizon is assumed to be some fraction, $\gamma$ ,of the horizon mass at that epoch.
The mass  is therefore given by \cite{Green:1997sz}
\begin{equation}
M_{\mathrm{PBH}}(k)=\gamma \frac{4\pi }{3}\rho H^{-3}\bigg|_{k=aH} ~.
\end{equation}
where $\rho$ is the energy density, and $a$ and $H$ correspond to the scale factor and the Hubble expansion rate of the Universe respectively
at the PBH formation epoch. Here $k=aH$ is the comoving scale of the mode associated with the density perturbation.
In our calculations, we assume $\gamma=0.2$ \cite{Carr:1975qj}. Since, in our study we study the PBH formation in a
radiation-dominated era, all the above quantities are defined in the radiation era.
So at the formation time, the PBHs mass is related to the comoving scale, $k$ via 
\begin{equation}
M_{\mathrm{PBH}}(k)\sim 5\times 10^{15}\mathrm{g}~\left( \frac{g_{\star,0}}{g_{\star,i}}\right)^{\frac{1}{6}} \left(\frac{10^{15}~\mathrm{Mpc}^{-1}}{k} \right)^{2} 
\label{eq:masskrel}
\end{equation}
where $g_{\star,0}$ and $g_{\star,i}$ are the relativistic degree of freedom associated with the energy density at present and PBH formation epoch. 
When PBH forms, its initial mass fraction, $\beta(M_{\mathrm{PBH}})$, is defined as
\begin{equation}
\beta(M_{\mathrm{PBH}})=\frac{\rho_{\mathrm{PBH},i}}{\rho_{\mathrm{total},i}}
\end{equation}
where $\rho_{\mathrm{PBH},i}$ and $\rho_{\mathrm{total},i}$ denote the energy densities of PBHs and the total energy density of the universe respectively
at the time of PBH formation. Assuming, $\rho_{\mathrm{PBH},i}\propto a^{-3}$ and $\rho_{\mathrm{total},i}\propto a^{-4}$,  we can rewrite $\beta(M_{\mathrm{PBH}})$ in terms of the present values of respective quantities as 
\begin{equation}
\beta(M_{\mathrm{PBH}})=\frac{\Omega_{\mathrm{PBH},0}(M_{\mathrm{PBH}})}{\Omega^{\frac{3}{4}}_{r,0}~\gamma^{\frac{1}{2}}}\left( \frac{g_{\star,i}}{g_{\star,0}}\right)^{\frac{1}{4}} \left(\frac{M_{\mathrm{PBH}}}{M_{H0}} \right)^{\frac{1}{2}} 
\label{eq:betaob}
\end{equation}
where $\Omega_{r,0}=\rho_{r,0}/\rho_{\mathrm{crit},0}$ is the ratio of present radiation energy density to the critical energy, $\rho_{\mathrm{crit},0}$
and  $M_{H0}=\frac{4\pi }{3}\rho_{\mathrm{crit},0} H_{0}^{-3}$ is the present horizon mass.

The current density parameter for the PBHs which have not been evaporated yet is 
\begin{equation}
\Omega_{\mathrm{PBH},0}(M_{\mathrm{PBH}})=\frac{\rho_{\mathrm{PBH},0}(M_{\mathrm{PBH}})}{\rho_{\mathrm{crit},0}}~~.
\label{eq:PBH0}
\end{equation}
where $\rho_{\mathrm{PBH},0}$ is the present PBH energy density.
The quantity $f_{\mathrm{PBH}}$, defined as the fraction of current PBH mass density to the current Cold Dark Matter (CDM) density is given by 
\begin{equation}
f_{\mathrm{PBH}}=\frac{\Omega_{\mathrm{PBH},0}(M_{\mathrm{PBH}})}{\Omega_{\mathrm{CDM},0}}~~.
\label{eq:fPBH0}
\end{equation}
Using Eq. (\ref{eq:betaob}), (\ref{eq:PBH0}) and (\ref{eq:fPBH0}), we obtain
\begin{equation}
f_{\mathrm{PBH}}=\frac{\beta(M_{\mathrm{PBH}})\Omega^{\frac{3}{4}}_{r,0}\gamma^{\frac{1}{2}}}{\Omega_{\mathrm{CDM},0}}\left( \frac{g_{\star,i}}{g_{\star,0}}\right)^{-\frac{1}{4}} \left(\frac{M_{\mathrm{PBH}}}{M_{H0}} \right)^{-\frac{1}{2}} ~.
\label{eq:fPBH}
\end{equation}
The Press-Schechter theory \cite{Press:1973iz} can now be used to obtain the expression of $\beta(M_{\mathrm{PBH}})$:
\begin{equation}
\beta(M_{\mathrm{PBH}})= \mathrm{Erfc} \left( \frac{\delta_{c}}{\sqrt{2}\sigma(R)}\right) ~,
\label{eq:betath}
\end{equation}
where $\mathrm{Erfc}(y)=\frac{2}{\sqrt{\pi}}\int^{\infty}_{y} e^{-t^{2}} dt$ represent the complementary error function and $\delta_{c}$ is the critical value of density
perturbation required for the PBH formation. 
A simple theoretical calculation gives $\delta_{c}=\frac{1}{3}$ \cite{1975ApJ...201....1C} and numerical simulations obtained
$\delta_{c}=0.4-45$, see Refs. \cite{Harada:2013epa, Harada:2015yda,Kohri:2018qtx} and references therein.
We take $\delta_{c}=0.42$ for our calculations.
 Here $\sigma(R)$ is the mass variance which can be estimated at the horizon
crossing via \cite{2000cils.book.....L}
\begin{equation}
\sigma^{2}(R)=\int \tilde{W}^{2}(kR) \mathcal{P}_{\delta}(k)\frac{dk}{k}
\label{eq:massvari}
\end{equation}
where $P_{\delta}(k)$ and $\tilde{W}(kR)$ represent the matter power spectrum and the Fourier transform of the
window function respectively. Here, $w$ is equation of state of fluid, and is equal to $\frac{1}{3}$ for radiation-dominated epoch. In our calculation, we consider a Gaussian window function i.e., $\tilde{W}(kR)=\exp{(-k^{2}R^{2}/2)}$.
The matter power spectrum $P_{\delta}$ is related to $P_{\mathcal{R}}$ by the equation
\begin{equation}
\mathcal{P}_{\delta}(k)=4\left(\frac{1+w}{5+3w}\right)^{2} \mathcal{P}_{\mathcal{R}}(k)~.
\label{eq:densityPS}
\end{equation}
Therefore, using Eq. (\ref{eq:densityPS}), Eq.(\ref{eq:massvari}), and Eq.(\ref{eq:betath}) in Eq. (\ref{eq:fPBH}), we can calculate
the PBH dark matter fraction, $f_{\mathrm{PBH}}$, as a function of PBH mass.

We have used $h^{2}\Omega_{\mathrm{CDM},0}=0.11933, ~ h^{2}\Omega_{\mathrm{B},0}=0.02242$ and $h=0.6766$ \cite{Planck:2018vyg} in our numerical computations.

 \subsection{Scalar Induced Gravitational Waves }
 \label{Subsec2}
The scalar modes couple to the tensor modes at the second order in perturbation theory. At the time of PBH formation, the scalar
perturbations are enhanced giving rise to the possibility of significant second order tensor perturbations.
Thus, PBH production indirectly induce gravitational waves \cite{Mollerach:2003nq,Ananda:2006af,Baumann:2007zm,Saito:2008jc}.
The fraction of GW energy density per logarithmic $k$ interval to the total energy density is given by \cite{Baumann:2007zm,Kohri:2018awv}
 \begin{align}
 \Omega_{\mathrm{GW}}(\eta,k)=\frac{1}{24}\left( \frac{\mathrm{k}}{a(\eta)H(\eta)}\right)^{2} \overline{\mathcal{P}_h(\eta, \mathrm{k})}
 \label{eq:omegagw}
 \end{align}
where, $\overline{\mathcal{P}_h(\eta, \mathrm{k})}$ is the dimensionless power spectrum averaged over time. Assuming vanishing anisotropic stress and neglecting the non-Gaussianity in the primordial curvature power spectrum, we obtain
 \cite{Baumann:2007zm,Kohri:2018awv}
 \begin{align}
 \mathcal{P}_{h}(\eta,k) = 4\int\limits_{0}^{\infty} dv \int\limits_{|1-v|}^{|1+v|} du \left[ \frac{4v^{2}-(1+v^{2}-u^{2})^{2}}{4vu}\right]^{2} \nonumber \\ \times  ~[\mathrm{I}(v,u,x)]^{2}P_{\mathcal{R}}(kv)P_{\mathcal{R}}(ku)
 \label{eq:phuv}
 \end{align} 
where $u=| \textbf{k}-\tilde{\textbf{k}}|/\mathrm{k}$ and $v=\tilde{\mathrm{k}}/\mathrm{k}$ are dimensionless variables. In the above
equation, $x\equiv k\eta$ and the function $\mathrm{I}(v,u,x)$ is defined in Ref. \cite{Kohri:2018awv}.
Using Eq. (\ref{eq:omegagw}), the energy spectrum of induced gravitational waves $\Omega_{\mathrm{GW},0}(k)$ at the present time
can be obtained \cite{Bastero-Gil:2021fac}:
 \begin{equation}
 \Omega_{\mathrm{GW},0}(k)=0.39\left(\frac{g_{\star}}{106.75} \right)^{-\frac{1}{3}} \Omega_{r,0}~ \Omega_{\mathrm{GW}}(\eta_{c},k)
 \label{eq:omegagw0}
 \end{equation}
where $\eta_{c}$ is the conformal time when a perturbation is inside the horizon during radiation dominated era. In the above equation, $\Omega_{r,0} \approx 9\times 10^{-5} $ is the present radiation energy density, and $g_{\star} $ is the effective number of relativistic degree of freedom in the radiation dominated era. To calculate the gravitational wave energy density, $\Omega_{\mathrm{GW},0}$, as a function of frequency,
we express $k$ in terms of frequency, $f$, using
 \begin{equation}
 f=\frac{k}{2\pi}=1.5\times 10^{-15}\left( \frac{k}{1~ \rm{Mpc}^{-1}}\right)  \rm{Hz}.
 \label{eq:frkrel}
 \end{equation}
 \section{ Parametrisation of primordial curvature perturbations }
\label{Sec2}
Amplitude of the power spectrum at large scales $k \sim 0.05$ Mpc$^{-1}$ is measured from CMB to be $P_\mathcal{R} =2.1\times 10^{-9}$ \cite{Akrami:2018odb}.
PBHs formation at small scales however requires much larger amplitude, $P_\mathcal{R} (k)\sim \mathcal{O} (10^{-2})$.
Therefore, scale-invariant, $n_{s}=1$ and red-tilted, $n_{s}<1$ power spectrum will not be able to form PBHs.
In order to form PBHs, the curvature power spectrum needs to be enhanced by several orders of magnitudes at small scales.
In this work,  following Ref. \cite{Li:2018iwg}, we assume a phenomenological form of curvature power spectrum parametrised by the scalar
spectral index, it's running and also it's running of running:
\begin{equation}
\mathcal{P}_{\mathcal{R}}(k)=A_{s}\left( \frac{ k}{k_{p}}\right)^{n_{s}-1+\frac{\alpha_{s}}{2}  \log\left(  \frac{k}{k_{p}}\right) +\frac{\beta_{s}}{6}  \left(\log\left(  \frac{k}{k_{p}}\right)\right)^{2}}~~,
\label{eq:PS}
\end{equation}
where $k_{p}=0.05$ Mpc$^{-1}$ is the pivot scale. Here, $n_{s}$, $\alpha_{s} \equiv d n_{s}/d\ln k $ and $\beta_{s} \equiv d^{2} n_{s}/d\ln k^{2} $ correspond to scalar spectral index, it's running and also it's running of running.
This kind of parametrisation has also been discussed earlier in the literature \cite{Kosowsky:1995aa,Planck:2018jri,Green:2018akb}.
 For positive values of running parameters, i.e., $\alpha_{s}>0$, $\beta_{s}>0$, $\mathcal{P}_{\mathcal{R}}(k)$ can be enhanced at large $k$,
 and can, in principle, lead to PBH formation.
 From Eq.~(\ref{eq:masskrel}), it is also clear that larger (smaller) $k$ will produce  smaller (larger) mass PBHs.

For our estimation, we consider $\alpha_{s}$ and $\beta_{s}$ values obtained from Planck and BAO data \cite{Li:2018iwg} (see also, \cite{Planck:2018jri})
\begin{equation*}
\ln (10^{10} A_{s})=3.088\pm 0.023, \quad n_{s}= 0.9660\pm 0.0040~,
\vspace{-0.3cm}
\end{equation*}
\begin{equation}
\alpha_{s}=0.0077^{+0.0104}_{-0.0103}, \quad \beta_{s}= 0.019\pm 0.013~~.
\end{equation}
Assuming, $A_{s}= 2.19\times 10^{-9}, n_{s}=0.09660$, PBH formation at scale $k$ requires,
\begin{equation}
(20k)^{-0.034+\frac{\alpha_{s}}{2}  \log\left( 20k\right) +\frac{\beta_{s}}{6}  \left(\log\left(20k\right)\right)^{2}}\sim 4.56\times 10^{6}
\end{equation}
It is evident that there is a degeneracy between $\alpha_{s}$ and $\beta_{s}$ parameters for PBH formation at some scale $k$.
Note that the $\alpha_{s}$ and $\beta_{s}$ parameters are related to higher order slow-roll parameters \cite{Li:2018iwg}; therefore
they can in principle tell us about the inflationary models \cite{Drees:2011yz}. %
PBH formation in the context of an enhanced power spectrum has also been discussed in
\cite{Josan:2009qn,Drees:2011hb,Sasaki:2018dmp,Kohri:2018qtx}.
\begin{figure}[]
	\centering
\includegraphics[height=2.in,width=3.in]{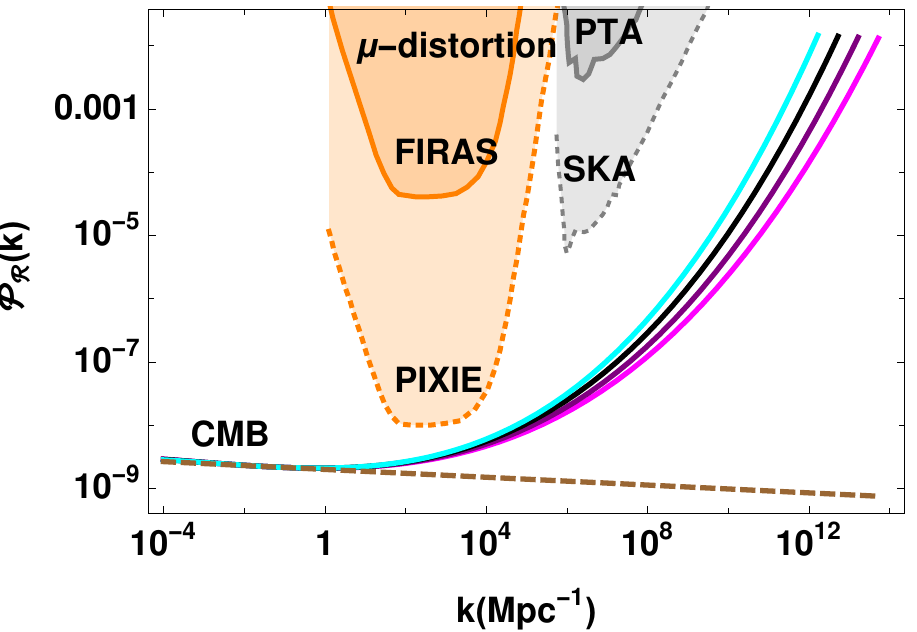}
\caption{The primordial curvature power spectrum as a function of the scale $k$ for different set of values of $\alpha_{s}$ and $\beta_{s}$. The constraints on $\mathcal{P}_{\mathcal{R}}(k)$ from $\mu-$distortion, PTA and SKA are also shown \cite{Chluba:2019kpb}.
} 
\label{fig:pk}	
\end{figure}
 \section{Results and discussions}
 \label{Sec3}
Following the basic formalism of PBH formation and the parametrisation of the curvature power spectrum,
we can now discuss the PBH production in the mass range $\sim 10^{17}$g$-10^{22}$g constituting the
the whole of DM density. 
We consider that the PBHs in the aforementioned mass range form when the corresponding modes re-enter the
horizon after inflation during a radiation-dominated era.

Before discussing the PBH formation, let us first show 
the primordial curvature power spectrum, see Fig. (\ref{fig:pk}), as a function of $k$ for different set of values of $\alpha_{s}$ and $\beta_{s}$. 
The magenta ($0.00862, 0.0017$), purple ($0.0083, 0.00193$), black ($0.00873,0.00216$) and cyan ($0.00844, 0.0025$) curves correspond to four benchmark values of $\alpha_{s}, \beta_{s}$. Here we see that $\mathcal{P}_{\mathcal{R}}(k)$ corresponding to benchmark values are consistent with the Planck measurement at small $k$ but the amplitude grows sharply to $\mathcal{P}_{\mathcal{R}}(k)\sim \mathcal{O}(10^{-2})$ at large $k$ allowing the formation of PBHs. 
The brown-dashed line corresponds to the power spectrum with $\alpha_{s}=0, \beta_{s}=0$.
Note that all the lines are consistent with the Planck measurement at small $k$.
 In the case of $\alpha_{s}=0, \beta_{s}=0$, the spectral index is red, i.e., $n_{s}<1$, and $\mathcal{P}_{\mathcal{R}}(k)$ decreases with increasing
 $k$ (thus being unable to produce PBHs).

\begin{figure}[]
	\centering
\includegraphics[height=2.in,width=3.in]{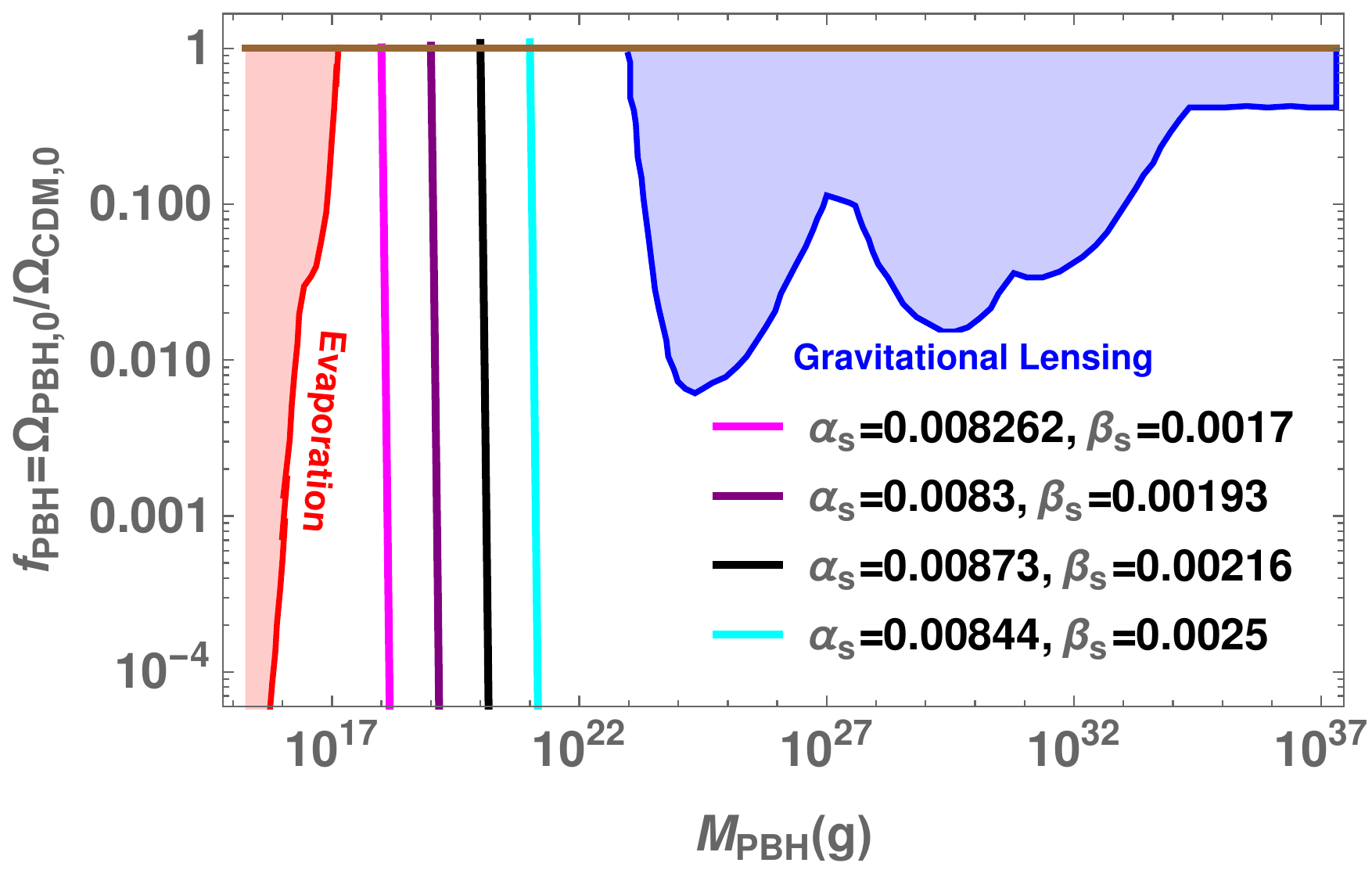}
\caption{The PBH fraction, $f_{\mathrm{PBH}}$, corresponding to four set of benchmark values of $\alpha_{s}$, $\beta_{s}$ showing that
the PBHs can indeed be the whole of DM in the asteroid mass region.
The constraints on $f_{\mathrm{PBH}}$ from evaporation and gravitational lensing \cite{Carr:2020xqk} are also shown.}
\label{fig:PBHfractmodel}		
 \end{figure}
Fig.~(\ref{fig:PBHfractmodel}) shows the PBH fraction, $f_{\mathrm{PBH}}$, as a function of the PBH mass, $M_{\mathrm{PBH}}$ corresponding to the
($\alpha_{s}$, $\beta_{s}$) values taken in Fig. (\ref{fig:pk}). The existing constraints on the PBH fraction as DM are also superimposed.
The magenta, purple, black and cyan curves show that $f_{\mathrm{PBH}}=1$ can indeed be obtained in the asteroid mass region, for example,
for the PBH masses $10^{18}$g, $10^{19}$g, $10^{20}$g and $10^{21}$g respectively.

\begin{figure}[]
	\centering
 \includegraphics[height=2.2in,width=3.2in]{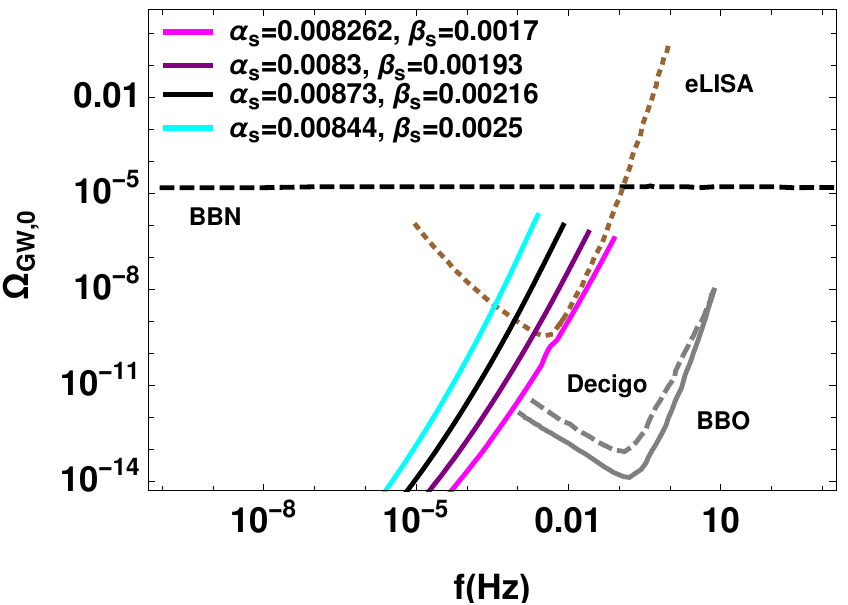}
 \caption{The scalar induced GW energy density as a function of the frequency for the four benchmark parameter sets.
 Sensitivity curves of some of the future GW detectors are also shown for comparison.}
	\label{fig:gw}	
 \end{figure}

 In Fig. \ref{fig:gw} we plot the present GW energy density as a function of the frequency again for the same four set of benchmark
 values of $\alpha_{s}$, $\beta_{s}$. The sensitivity curves for the present and future GW detectors are taken from \cite{Maggiore:1999vm,Assadullahi:2009jc,2009Natur.460..990A,Moore:2014lga,Inomata:2018epa,Mandic}. The constraint on the GW background from BBN, i.e., 
 $\Omega_{\mathrm{GW},0}h^{2}<5\times 10^{-6}$ is taken from  \cite{Maggiore:1999vm,LIGOScientific:2006zmq}. One can see from Fig.~\ref{fig:gw} that the peak GW frequency lies in the  range: $10^{-3}$Hz $-10^{-1}$ Hz. 
So the sensitivity curves suggest that the GWs can be detected by future detectors like eLISA, BBO and DECIGO. 

 \section{Summary}
 \label{Sec5}
PBHs in the asteroid mass window $~ 10^{17}- 10^{22}$g are currently unconstrained to constitute the whole of DM density of our universe. 
In this paper, we explore the possibility of formation of PBHs in this mass range considering a phenomenological parametrisation (in terms of
two parameters $\alpha_s$ and $\beta_s$)
of an enhanced curvature power spectrum.  
Using allowed values of $\alpha_s$ and $\beta_s$ from the Plank and BAO data, we have shown that one can indeed get full PBH DM abundance
in the said mass range.

Using the same set of $\alpha_s$ and $\beta_s$ parameters, we also calculated the scalar induced GW spectrum
and showed that they fall in the sensitivity region of future GW detectors like eLISA, BBO and DECIGO.

It would be nice to understand if GW signal, as indirect evidence of large PBH abundance in the asteroid mass region, have some characteristic features that can be used to distinguish it from GWs generated by other mechanisms, e.g. supermassive black holes. 
It would also be interesting to check what kind of  inflationary models can produce such a large power spectrum at small scales
required for significant PBH production. We leave these questions for future work. 
\section{Acknowledgement}
DG  and AKM acknowledge support through the Ramanujan Fellowship and MATRICS Grant of the Department of Science and Technology,
Government of India. We thank Arka Banerjee and Susmita Adhikari for valuable comments.  AKM would
also like to thank Richa Arya for fruitful suggestions and helping in the \textit{Mathematica} program.

\bibliographystyle{utphys}
\bibliography{Draft_Final}
\end{document}